\documentclass[final,1p,times]{elsarticle} 
\usepackage{graphicx} 
\usepackage{amssymb} 
\usepackage{amsthm} 
\usepackage{lineno} 

\usepackage{bm}
\usepackage{amsmath}
\usepackage{color}

\newcommand{\bald}[1]{{\bf #1}}

\usepackage{dcolumn}

\journal{Nuclear Physics A} 
\begin{document} 

\begin{frontmatter} 


\title{The sound generated by a fast parton in the quark-gluon plasma\\ 
is a {\em crescendo}}

\author{R. B. Neufeld and B. M\"uller}

\address{Department of Physics, Duke University, Durham, NC 27708, USA}

\begin{abstract} 
The total energy deposited into the medium per unit length by a fast parton traversing a quark-gluon plasma is calculated.  We take the medium excitation due to collisions to be given by the well known expression for the collisional drag force.  The parton's radiative energy loss contributes to the energy deposition because each radiated gluon acts as an additional source of collisional energy loss in the medium.  In our model, this leads to a length dependence on the differential energy loss due to the interactions of radiated gluons with the medium.  The final result, which is a sum of the primary and the secondary contributions, is then treated as the coefficient of a local hydrodynamic source term.  Results are presented for energy density wave induced by two fast, back-to-back partons created in an initial hard interaction.
\end{abstract} 

\end{frontmatter} 


The question of how the quark-gluon plasma responds to fast partons as they propagate through it has gained significance in light of measurements of hadron correlation functions that are consistent with a conical emission pattern \cite{Adams:2005ph,Adler:2005ee,Ulery:2007zb,Adare:2008cq}.  These measurements suggest that fast partons may generate collective conical disturbances in the medium.  Theoretical studies of how the quark-gluon plasma responds to fast partons are relatively recent \cite{Machcone,Renk:2005si,Betz:2008js,Neufeld:2008fi,Neufeld:2008hs,Neufeld:2008dx}. One of the primary challenges in this investigation is to determine the energy deposited into the medium per unit length by an energetic parton.  As it propagates through the medium, the fast parton loses energy through collisions and medium induced radiation.  The energy deposited into the medium per unit length is then the sum of the collisional energy loss of the primary parton and of the radiated gluons.  The energy deposited into the medium per unit length is the sum of the collisional energy loss of the primary parton and of the radiated gluons.

In what follows, we calculate the total energy deposited into the medium per unit length by a fast parton traversing a quark-gluon plasma \cite{Neufeld:2009ep}.  We start by considering the collisional energy lost per unit length by the fast parton for which we will use the familiar expression \cite{Thoma:1991ea}
\begin{equation}\label{dE_dx_c}
\left(\frac{dE}{d \xi}\right)_C = \frac{\alpha_s C_2 \, m_{D}^2}{2}  \ln \frac{2 \sqrt{E_p T}}{m_D}
\end{equation}
where the subscript $C$ denotes collisional energy loss and $\xi$ is the path-length traveled by the source parton.  Also, $\alpha_s = g^2/4\pi$ is the strong coupling, $m_{D}$ is the Debye mass of the medium, which we take to be given by $ m_{D} =  gT$, $E_p$ is the energy of the fast parton, $T$ is the temperature of the medium, and $C_2$ is the eigenvalue of the quadratic Casimir operator of the color charge of the source parton, which is 4/3 if the fast parton is a quark, and 3 for a gluon.  

We next calculate the energy gained by the medium due to gluons radiated by the fast parton.  We begin by deriving a partial differential equation through which one can determine the spectrum of radiated gluons in the medium, defined here as $f(\omega,\xi)$.  $f(\omega,\xi)$ will in general differ from the spectrum of gluons emitted by the fast parton, $d I /d \omega$, because gluons, once emitted, lose energy in the medium due to collisions until they become part of the thermal bath.  Specifically, as a gluon with energy $\omega$ travels from $\xi$ to $\xi + \Delta \xi$, it loses collisional energy $\epsilon(\omega)\, \Delta \xi$, where $\epsilon(\omega)$ is obtained from (\ref{dE_dx_c}) to be
\begin{equation}
\epsilon(\omega) = \frac{3}{2} \alpha_s \, m_{D}^2 \ln \frac{2 \sqrt{\omega T}}{m_D}.
\end{equation}
It follows that in order to find a gluon with energy $\omega$ at position $\xi + \Delta \xi$, there must be a gluon with energy $\omega + \epsilon(\omega)\, \Delta \xi$ at position $\xi$.  Furthermore, we require $f \, d \omega$, that is, the total number of gluons, to be invariant, which leads to the following relation
\begin{equation}\label{crude}
f(\omega,\xi + \Delta \xi) = f(\omega + \epsilon(\omega)\, \Delta \xi,\xi) \left(1 + \frac{\partial \epsilon(\omega)}{\partial \omega} \Delta \xi \right).
\end{equation}
Finally, as the fast parton moves from $\xi$ to $\xi + \Delta \xi$, it will emit additional gluons, $\Delta \xi \times d I/d\omega d\xi$, which add to $f(\omega,\xi+\Delta\xi)$, so that in the limit of $\Delta \xi \rightarrow 0$ the evolution equation for $f(\omega,\xi)$ takes the form
\begin{equation}\label{diff_e_q}
\frac{\partial}{\partial \xi}f(\omega,\xi) - \frac{\partial}{\partial \omega} \left[\epsilon(\omega)\, f(\omega,\xi)\right] = \frac{d I}{d \omega d \xi}(\omega,\xi).
\end{equation}
To determine $f$, it is necessary to specify $dI/d\omega d\xi$, for which we choose the spectrum calculated by Salgado and Wiedemann in the multiple soft scattering approximation \cite{Salgado:2003gb}
\begin{equation}\label{specchoice}
\frac{d I}{d \omega d \xi} = - \frac{\sqrt{\hat{q}} \, \alpha_s C_2}{\pi} \text{Re} \frac{(1 + i)\tan \left[(1+i)\sqrt{\frac{\hat{q}}{\omega}} \frac{\xi}{2}\right]}{\omega^{3/2}}
\end{equation}
where we use the following expression for the jet quenching parameter \cite{Baier:2008zz}:
\begin{equation}\label{qhat}
\hat{q} = 2 \alpha_s C_2 m_D^2 T \ln \frac{2 \sqrt{E_p T}}{m_D}.
\end{equation}

We make the distinction that gluons with energy $\omega > \bar{\omega} \equiv 2 T$ are sources of collisional energy loss, while those with $\omega < \bar{\omega}$ immediately become part of the medium.  For $\omega > \bar{\omega}$ we solve for $f(\omega,\xi)$ numerically from equation (\ref{diff_e_q}) for a primary fast quark using the parameters: $\alpha_s = 1/\pi$, $T = 300$ MeV, and $E_p = 50$ GeV.  The total energy deposited into the medium by the secondary gluons per unit length takes the form
\begin{equation}\label{dE_dx_r}
\begin{split}
\left(\frac{dE}{d \xi}\right)_R = \int_{\bar{\omega}}^{\omega_{\rm max}} d \omega \, &\epsilon(\omega) \, f(\omega,\xi) + \int_{0}^{\bar{\omega}} d \omega \, \omega \frac{d I}{d \omega d \xi} + \bar{\omega} \, f(\bar{\omega},\xi)\, \epsilon(\bar{\omega}),
\end{split}
\end{equation}
where we set $\omega_{\rm max} = E_p/2$.  The final term in equation (\ref{dE_dx_r}) accounts for energetic gluons which have lost enough energy to become a part of the medium and deposit their entire remaining energy.   We multiply the spectrum (\ref{specchoice}) by a factor of $1 - (\omega/\omega_{\rm max})^8$ to ensure it goes to zero at $\omega = \omega_{\rm max}$ when solving (\ref{diff_e_q}).

\begin{figure*}
\centerline{
\includegraphics[width = 0.5\linewidth]{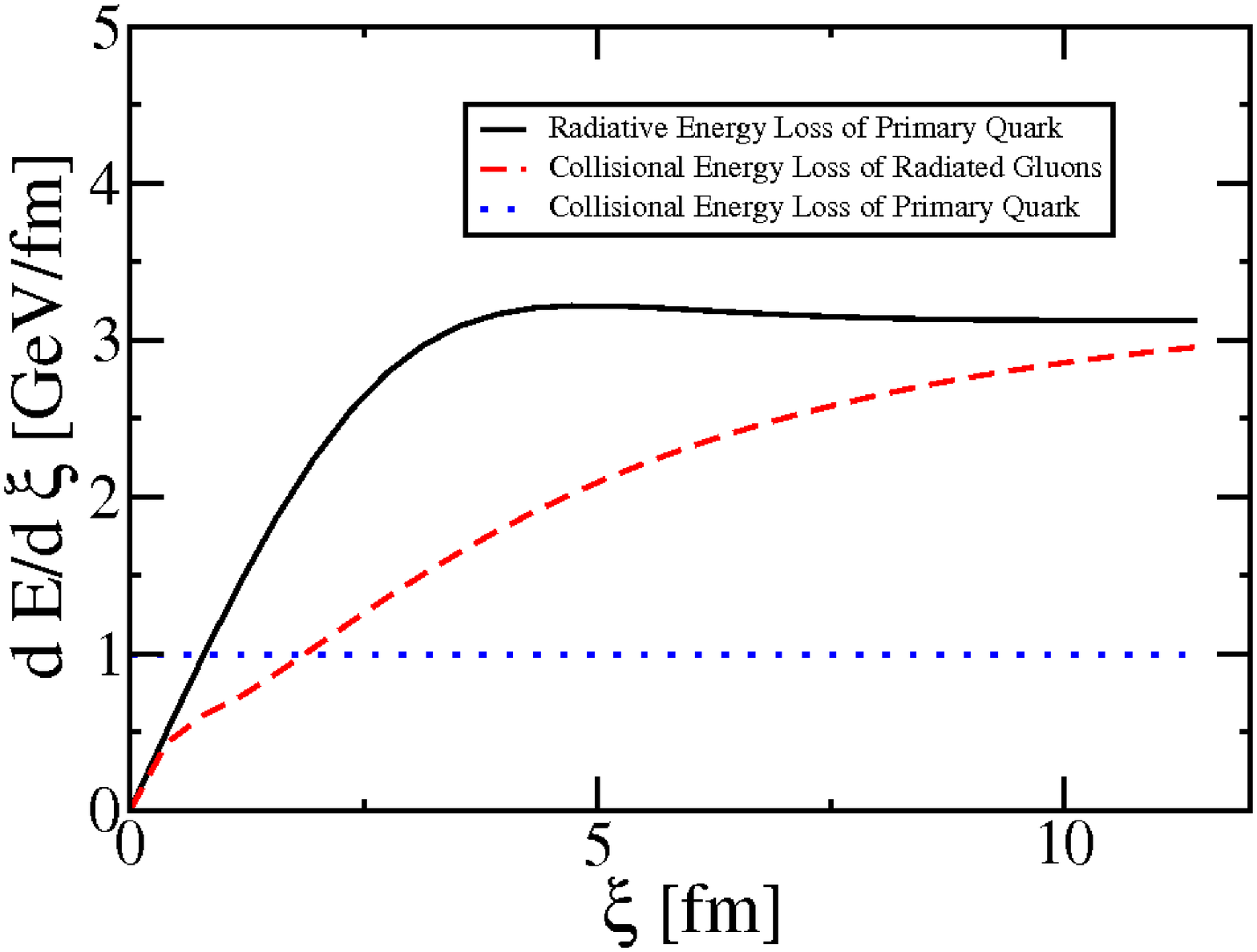}
}
\caption{The result for (\ref{dE_dx_r}) is shown as the dashed line in Fig.~\ref{total_energy} for the parameters listed above, along with the differential collisional energy loss of the primary parton (dotted line) and its differential energy loss to radiation (smooth line).  The differential energy deposition into the medium is the sum of the dotted and dashed lines. 
}
\label{total_energy}
\end{figure*}

The result for (\ref{dE_dx_r}) is shown as the dashed line in Fig.~\ref{total_energy} for the parameters listed above, along with the differential collisional energy loss of the primary parton (dotted line) and its differential energy loss to radiation (smooth line).  
The energy deposited by the radiation into the medium per unit length has an approximately linear growth in pathlength, which results from the steady increase in the number of gluons that deposit energy in collisions.  The total energy deposition into the medium per unit length, or time, is the sum of (\ref{dE_dx_c}) and (\ref{dE_dx_r}), which we write as
\begin{equation}\label{tot_diff_loss}
\frac{d E}{d t} = \left(\frac{dE}{d t}\right)_C + \left(\frac{dE}{d t}\right)_R
\end{equation}
We make the connection to the hydrodynamic response of the medium by treating the fast parton as a point source of energy and momentum deposition in the medium, with velocity $\bald{u}$ and energy $E_p$.  For a relativistic point source, the hydrodynamic source term, denoted as $J^\nu$, takes the following form
\begin{equation}\label{simp_source}
\begin{split}
J^\nu(x) &= \frac{dE}{dt}\delta(\bald{x} - \bald{u} t)\left(1,\bald{u}\right)
\end{split}
\end{equation}
where $dE/dt$ is given by (\ref{tot_diff_loss}).  The calculation is made more tractable by fitting the result of (\ref{dE_dx_r}) to a linear function of time, from which we find that for $t < 6$ fm
\begin{equation}
\left(\frac{dE}{d t}\right)_R \approx 0.3 \frac{\text{GeV}}{\text{fm}} + 0.364 \, t \frac{\text{GeV}}{\text{fm}^2} .
\end{equation}

The source term couples to the hydrodynamic equations for the medium, $\partial_\mu T^{\mu \nu} = J^{\nu}$, where $T^{\mu \nu}$ is the energy-momentum tensor.  We consider the source to be coupled to an otherwise static medium at temperature $T$ and that the energy and momentum density deposited by the fast parton is small compared to the equilibrium energy density of the medium so that the hydrodynamical equations can be linearized.  We are here interested in calculating the energy density perturbation excited in the medium, denoted as $\delta \varepsilon \equiv \delta T^{00}$.  The equations of motion for a medium coupled to a source in linearized hydrodynamics are discussed in several places (for instance, \cite{Neufeld:2008dx}).  
\begin{figure*}
\centerline{
\includegraphics[width = 0.95\linewidth]{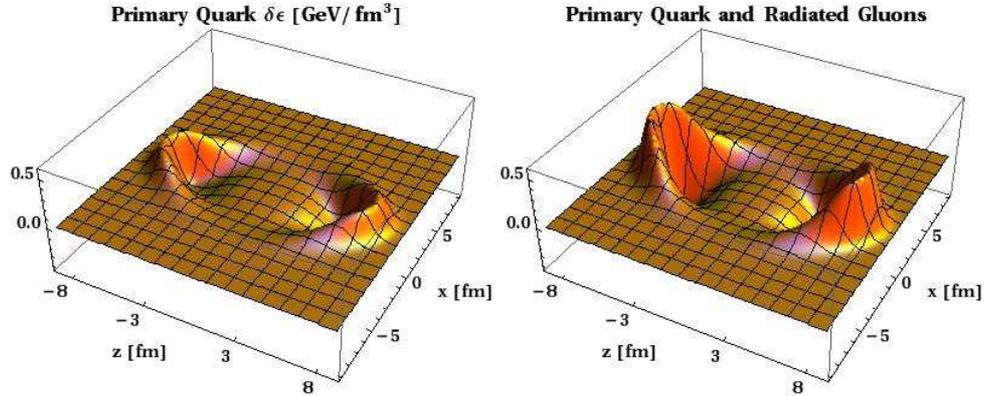}
}
\caption{Result for the energy density wave excited by back-to-back quarks propagating along the $\hat{z}$ axis.  The plots are shown in the $x$-$z$ plane, however, the results are cylindrically symmetric about the $\hat{z}$ axis. In the left figure, only the energy lost in collisions by the primary parton is considered; the right figure also includes the collisional energy loss of the radiated gluons.
}
\label{mach_wave}
\end{figure*}
We calculate the medium response at a time $t = 8$ fm for back to back quarks (each with energy $E_p = 50$ GeV) which are created at $t = 0$ and $\bald{x} = 0$, and propagate along the $\hat{z}$ axis until $t = 6$ fm.  The result is shown in Fig.~\ref{mach_wave}, for a medium with the same parameters used above as well as $\eta/s = 0.2$ and $c_s/c = 0.57$.  The plot shows the energy density wave excited by the source quarks in the $x$-$z$ plane, however, the results are cylindrically symmetric about the $\hat{z}$ axis.  The Mach cone formation is visible in the figure.  The radiative contribution to the source term exhibits a linear growth with time, which is reflected in the shape of the resulting energy density wave.


\section*{Acknowledgments}
This work was supported in part by the U.~S.~Department of Energy under grant DE-FG02-05ER41367. RBN thanks the organizers of the QM 2009 conference for local support. We acknowledge valuable discussions with A.~Majumder and draw attention to recent work by Qin {\em et al.} \cite{Qin:2009uh}, which addressed the same problem.

\end{document}